\documentclass[useAMS,usenatbib]{mn2e}
\usepackage{multirow}
\usepackage{epsfig}
\usepackage{subfigure}

\makeatletter
\let\jnl@style=\rm
\def\ref@jnl#1{{\jnl@style#1}}

\def\aj{\ref@jnl{AJ}}                   
\def\araa{\ref@jnl{ARA\&A}}             
\def\apj{\ref@jnl{ApJ}}                 
\def\apjl{\ref@jnl{ApJ}}                
\def\apjs{\ref@jnl{ApJS}}               
\def\ao{\ref@jnl{Appl.~Opt.}}           
\def\apss{\ref@jnl{Ap\&SS}}             
\def\aap{\ref@jnl{A\&A}}                
\def\aapr{\ref@jnl{A\&A~Rev.}}          
\def\aaps{\ref@jnl{A\&AS}}              
\def\azh{\ref@jnl{AZh}}                 
\def\baas{\ref@jnl{BAAS}}               
\def\jrasc{\ref@jnl{JRASC}}             
\def\memras{\ref@jnl{MmRAS}}            
\def\mnras{\ref@jnl{MNRAS}}             
\def\pra{\ref@jnl{Phys.~Rev.~A}}        
\def\prb{\ref@jnl{Phys.~Rev.~B}}        
\def\prc{\ref@jnl{Phys.~Rev.~C}}        
\def\prd{\ref@jnl{Phys.~Rev.~D}}        
\def\pre{\ref@jnl{Phys.~Rev.~E}}        
\def\prl{\ref@jnl{Phys.~Rev.~Lett.}}    
\def\pasp{\ref@jnl{PASP}}               
\def\pasj{\ref@jnl{PASJ}}               
\def\qjras{\ref@jnl{QJRAS}}             
\def\skytel{\ref@jnl{S\&T}}             
\def\solphys{\ref@jnl{Sol.~Phys.}}      
\def\sovast{\ref@jnl{Soviet~Ast.}}      
\def\ssr{\ref@jnl{Space~Sci.~Rev.}}     
\def\zap{\ref@jnl{ZAp}}                 
\def\nat{\ref@jnl{Nature}}              
\def\iaucirc{\ref@jnl{IAU~Circ.}}       
\def\aplett{\ref@jnl{Astrophys.~Lett.}} 
\def\apspr{\ref@jnl{Astrophys.~Space~Phys.~Res.}}
\def\bain{\ref@jnl{Bull.~Astron.~Inst.~Netherlands}}
\def\fcp{\ref@jnl{Fund.~Cosmic~Phys.}}  
\def\gca{\ref@jnl{Geochim.~Cosmochim.~Acta}}   
\def\grl{\ref@jnl{Geophys.~Res.~Lett.}} 
\def\jcp{\ref@jnl{J.~Chem.~Phys.}}      
\def\jgr{\ref@jnl{J.~Geophys.~Res.}}    
\def\jqsrt{\ref@jnl{J.~Quant.~Spec.~Radiat.~Transf.}}
\def\memsai{\ref@jnl{Mem.~Soc.~Astron.~Italiana}}
\def\nphysa{\ref@jnl{Nucl.~Phys.~A}}   
\def\physrep{\ref@jnl{Phys.~Rep.}}   
\def\physscr{\ref@jnl{Phys.~Scr}}   
\def\planss{\ref@jnl{Planet.~Space~Sci.}}   
\def\procspie{\ref@jnl{Proc.~SPIE}}   

\makeatother

\title[A multiwavelength map of the nuclear region of NGC~7582]{A multiwavelength map of the nuclear region of NGC~7582}

\author[Stefano Bianchi, Marco Chiaberge, Enrico Piconcelli, Matteo Guainazzi]{Stefano Bianchi$^1$\thanks{E-mail: Stefano.Bianchi@sciops.esa.int (SB)}, Marco Chiaberge$^2$, Enrico Piconcelli$^3$, Matteo Guainazzi$^1$\\
$^1$XMM-Newton Science Operations Center, European Space Astronomy Center, ESA, Apartado 50727, E-28080 Madrid, Spain\\
$^2$Space Telescope Science Institute, 3700 San Martin Drive, Baltimore, MD 21218\\
$^3$Osservatorio Astronomico di Roma (INAF), Via Frascati 33, I-00040 MontePorzio Catone, Italy\\
}

\begin{document}

\pagerange{\pageref{firstpage}--\pageref{lastpage}} \pubyear{2004}

\maketitle

\label{firstpage}

\begin{abstract}
We performed a detailed analysis of the \textit{Chandra} and \textit{HST} images of the Seyfert 2 galaxy, NGC~7582. The dust lane, as mapped by the \textit{HST} NIR and optical images, strongly influences the morphology of the soft X-ray emission, being spatially coincident with excess of X-ray absorption. Two `hot spots', i.e. regions where emission from higher ionization stages of O and Ne is stronger, are observed in the soft X-ray image. They can be tracers of variations of the ionization parameter, even if, at least for one of them, absorption may be the dominant effect. The positions of the `hot spots' suggest that they are not affected by the star-formation regions observed with \textit{HST}, being located far away from them. Therefore, the starburst does not seem to play an important role in the photoionization of the soft X-ray emitting gas. The resulting picture is consistent with modified unification scenarios, where the Compton-thick torus coexists with a large-scale Compton-thin material associated with the dust lane and circumnuclear gas is photoionized by the AGN along torus-free lines of sight.

\end{abstract}


\section{Introduction}

The origin of the soft X-ray emission in obscured Active Galactic Nuclei (AGN) is still under debate, but important progresses have been made in the last few years. The first breakthrough was represented by high resolution spectra made available thanks to the gratings aboard \textit{Chandra} and XMM-\textit{Newton}. The `soft excess' observed in CCD spectra was found to be due to the blending of bright emission lines, mainly from He- and H-like transitions of light metals and L
transitions of Fe, with low or no continuum, in a few bright objects \citep[e.g.][]{sako00b,Sambruna01b,kin02,brink02,schurch04,bianchi05b,pp05}. Spectral diagnostic tools agree that the observed lines should be produced in a gas photoionized by the AGN, with little contribution from any collisionally ionized plasma. These results have been recently confirmed to be common in a large catalog of Seyfert 2 galaxies (Guainazzi \& Bianchi, in preparation).  A second breakthrough was made possible thanks to the unrivaled spatial resolution of \textit{Chandra}. The soft X-ray emission of Seyfert 2 galaxies appears to be strongly correlated with that of the Narrow Line Region (NLR), as mapped by the [{O\,\textsc{iii}}] $\lambda 5007$ \textit{HST} images \citep[e.g.][]{yws01,bianchi06,levenson06}. Since the NLR is also believed to be a gas photoionized by the AGN, it was shown that a very simple model where the soft X-ray emission and the NLR emission are produced in the same material is possible \citep{bianchi06}. However, many relevant issues are still open. Is the soft X-ray emitting gas homogeneous? Which is its relation with the AGN environment (i.e. dust lanes, X-ray absorbing clouds)? Are there important sources of photoionization, like shocks \citep{king05} and starburst \citep{gd01}, other than the central AGN?

NGC~7582 is a nearby \citep[z=0.00525:][]{devac91} and bright Seyfert 2 galaxy, with no evidence of a Hidden Broad Liner Region (HBLR) in polarized light \citep{hlb97}. It shows a fairly extended and well-defined [{O\,\textsc{iii}}] cone \citep{sbb91}, along with a strong star-formation activity in the nuclear region \citep[see e.g.][]{wg06}. Though being observed by all the major X-ray satellites \citep{war93,turner97}, its very complex X-ray spectrum has been revealed only with BeppoSAX \citep{turn00} and XMM-\textit{Newton} (Piconcelli et al., in preparation). These authors have found that the source is obscured by at least two different materials. Moreover, the RGS high-resolution spectrum revealed for the first time the nature of its soft excess, completely dominated by several strong emission lines. The \textit{Chandra} image also shows that the soft X-ray emission is extended and contribution from starburst activities may be important \citep{dong2004}.

In this paper, we present a detailed comparison between the \textit{Chandra} and the \textit{HST} images of NGC~7582, in order to get insights on the properties of the extended soft X-ray emitting region and the role and characteristics of the obscuring materials in the circumnuclear region. 

\section{\label{obs}Observations and data reduction}

NGC~7582 was observed twice by \textit{Chandra}, in two consecutive ACIS \citep{acis} observations in October 2000 (obsid 436 and 2319), presented by \citet{dong2004}. We reduced data used in this paper with the Chandra Interactive Analysis of Observations (\textsc{ciao}) 3.3.0.1 and the Chandra Calibration Database (\textsc{caldb}) 3.2.1, adopting standard procedures. 
A flare in the background lightcurve is apparent in the middle of observation 436, but its peak is less than 10\% of the mean source count-rate, so not affecting our analysis. Images were corrected for known aspect offsets, reaching a nominal astrometric accuracy of 0.6 arcsec (at the 90\% confidence level). 
Since the two observations were performed very close in time and no significant variability is apparent between the two lightcurves, we decided to merge the two datasets using the \texttt{merge\_all} script, after having verified that any position difference between the two were smaller than 1 pixel ($\simeq0.49$ arcsec). The total exposure time for the merged dataset is 19 ks. We performed our analysis of the \textit{Chandra} image in the 0.3-10 keV band. In the following, we will refer to `hard X-rays' whenever $E>3$ keV.

\textit{HST} observations were retrieved from the MAST (Multimission Archive at STScI) and processed through the standard on-the-fly reprocessing system. NGC~7582 was observed with WFPC2 and F606W filter (average wavelength 5958.6 \AA, width 1579.0 \AA) on 06/15/1995 as part of  program GO~5479 (500s exposure time) and, with the same configuration, on 07/24/2001 as part of GO~8597 (560s exposure time).  The object was also imaged in the near-IR with NICMOS-NIC2 and the F160W filter (wavelength range 1.4-1.8 $\mu$m, central wavelength 1.55 $\mu$m) on 09/16/1997 (GO~7330). The central regions of the galaxy were imaged with WFPC2 using the `planetary camera', for which the projected pixel size is 0.0455" and the field of view is 36"x36". The pixel scale for NICMOS/NIC2 is 0.075", and the field of view is 19.2"x19.2". The FWHM of the PSF is $\sim 0.08"$ and $\sim 0.15"$ for the optical and IR images, respectively. The images were combined to remove cosmic ray, rectified and aligned using the \texttt{multidrizzle} task, available as part of the Pyraf/STSDAS data reduction package.

The cosmological parameters used throughout this paper are $H_0=70$ km s$^{-1}$ Mpc$^{-1}$, $\Lambda_0=0.7$ and $q_0=-0.5$.

\section{Analysis}
\label{analysis}
\begin{figure*}
\begin{center}
\epsfig{file=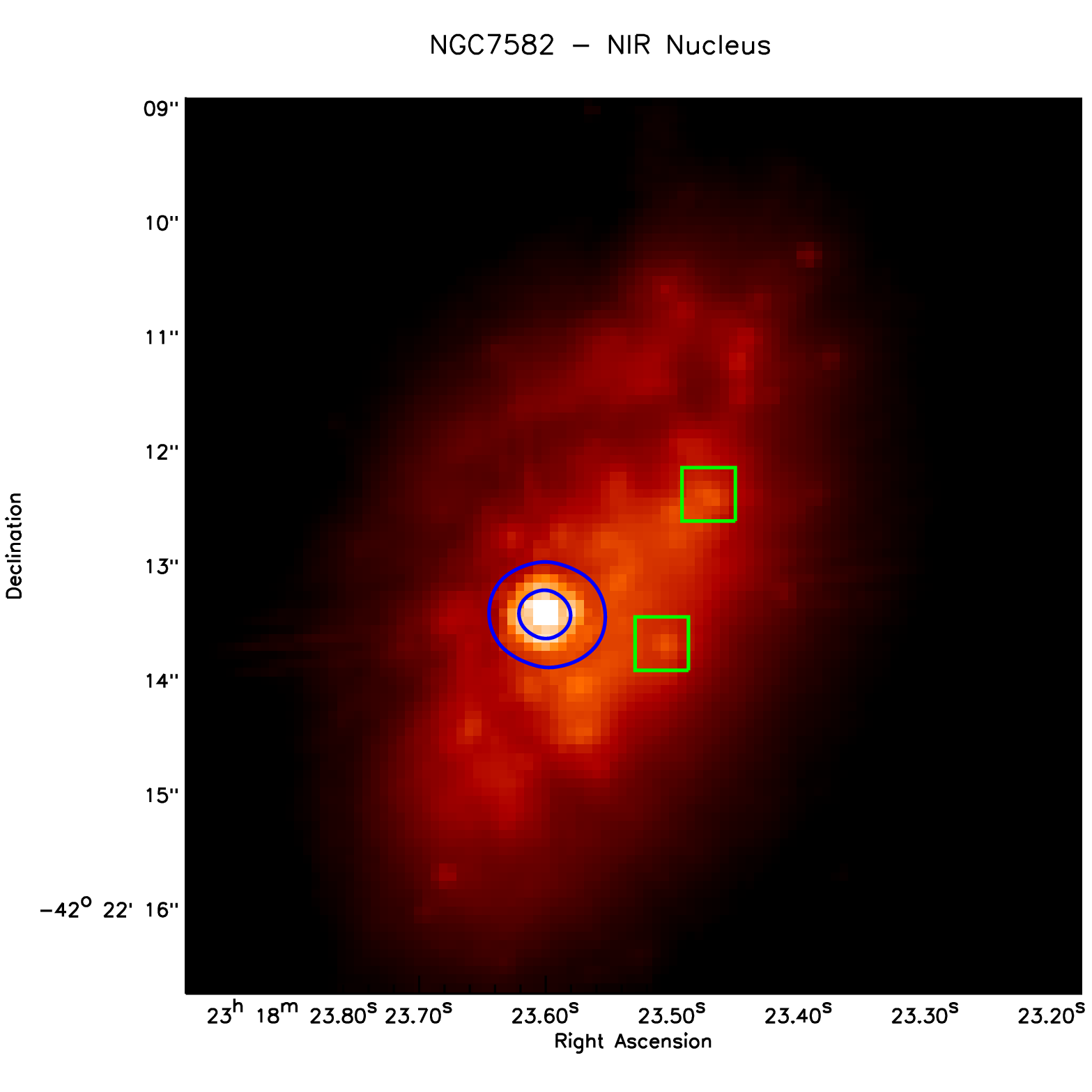, width=8.2cm}
\epsfig{file=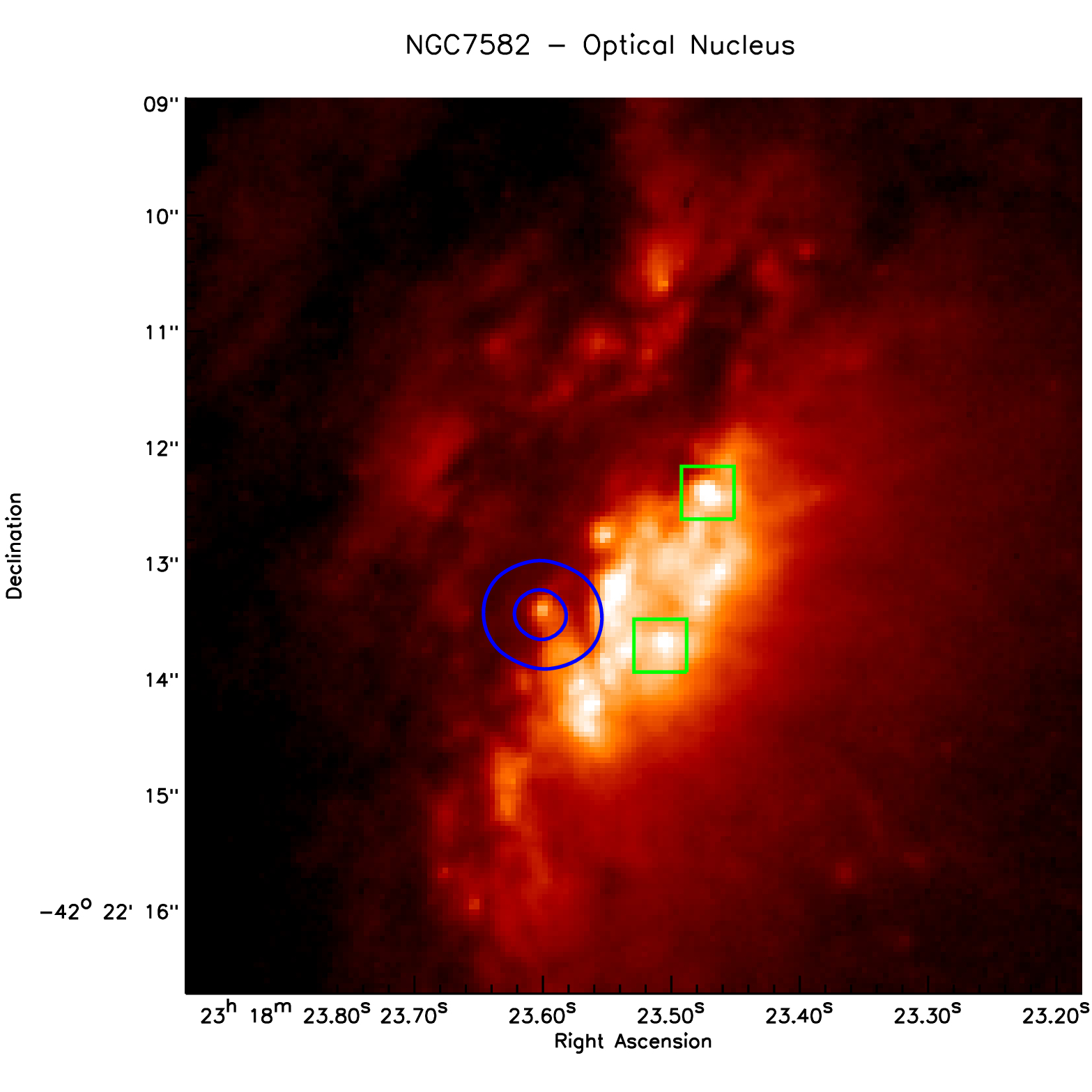, width=8.2cm}
\end{center}
\caption{\label{nucleus}NIR (\textit{left}) and optical (\textit{right}) \textit{HST} image of the nuclear region of NGC~7582. The \textit{Chandra} hard X-ray ($>3$ keV) contour plots are superimposed (in blue: they refer to the 90 and the 60\% of the peak), The star-formation regions used for the alignment of the two \textit{HST} images are highlighted by green boxes. North is up, east to the left.}
\end{figure*}

The \textit{HST} NIR image of NGC~7582 is dominated by a very bright unresolved nucleus (see Fig. \ref{nucleus}, left panel). We performed aperture photometry of the nucleus measuring the flux inside a radius of 4 pixels, corresponding to 0.3 arcsec. We measured the background at 4 pixels from the center, in an annulus of 1 pixel width. The largest source of error on the photometry is due to the estimate of the background, measured in a region where it is not constant. The near IR flux of the nucleus is $F_{1.6\mu}=(2.0\pm0.3)\times10^{-25}$ erg cm$^{-2}$ s$^{-1}$ hz$^{-1}$, i.e. a luminosity of $L_{1.6\mu}\simeq2.2\times10^{42}$ erg s$^{-1}$ at the distance of the source, very well in the range of Seyfert 2s and in agreement with its hard X-ray flux \citep[see e.g.][]{quillen01}. West of the nucleus, several star-formation regions are clearly visible. Some of them are easily detected also in the \textit{HST} optical images, where no strong nuclear emission is apparent. We therefore aligned the IR and optical images by means of the common star-formation regions, similarly to what done by \citet{wg06}. This procedure led to the identification of the optical nucleus, much dimmer than the IR counterpart. Finally, both \textit{HST} images were aligned with the hard X-ray \textit{Chandra} image, dominated by the nuclear unresolved emission, leading to a set of images in three bands with a consistent coordinate system (see Fig. \ref{nucleus}).

This identification of the optical nucleus is strengthened by the variation it shows between the two \textit{HST} observations, taken 6 years apart. Its flux at 5958.6 \AA\ (taken from a region with a 3 pixel radius) raised from 1.3 to $2\times10^{-28}$ erg cm$^{-2}$ s$^{-1}$ hz$^{-1}$, a variation of almost 60\% (the error on each measure is around 5\%). Conversely, for any other compact component (i.e. star formation regions) the flux is stable between the two observations, within the errors. The nucleus is unresolved, both in the optical and in the IR. The optical resolution of HST (i.e. $\sim 0.08$ arcsec) sets an upper limit on its physical size of 8 pc, at the distance of NGC~7582. 

\begin{figure*}
\begin{center}
\epsfig{file=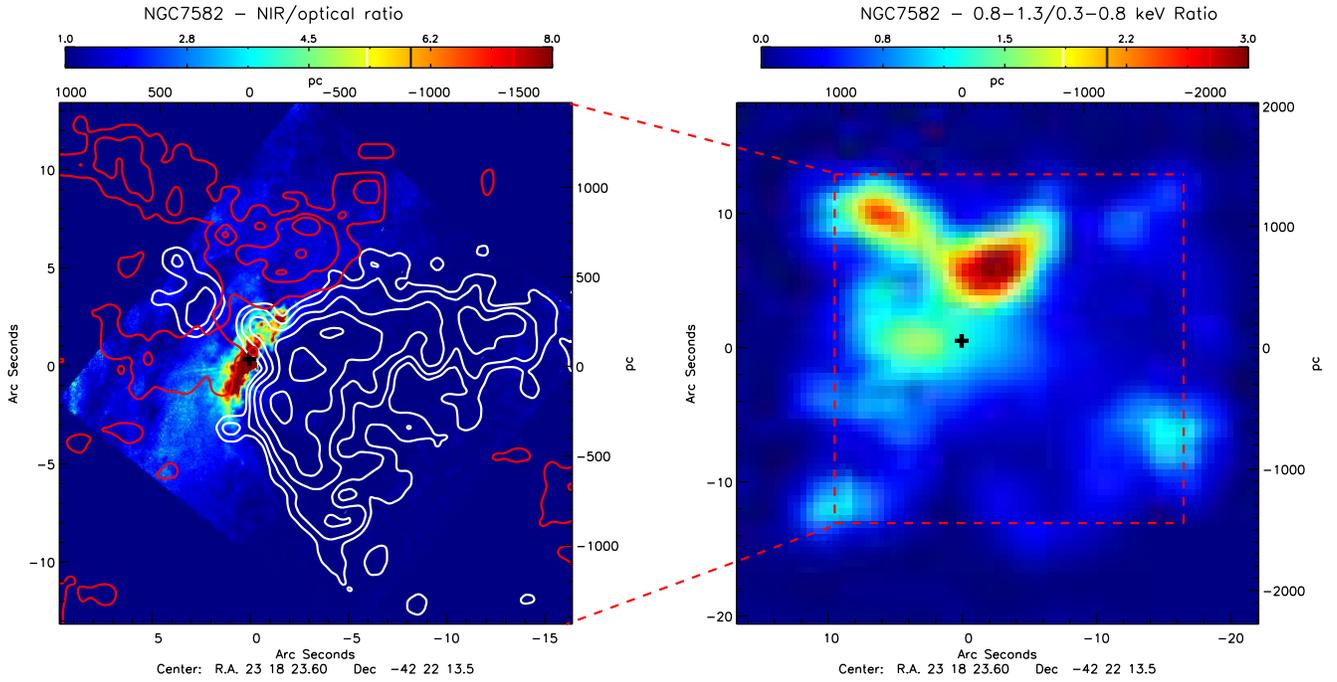, width=18.5cm}
\end{center}
\caption{\label{dustmap}\textit{Left:} \textit{HST} NIR to optical ratio, mapping the amount of dust in the circumnuclear region of NGC~7582. The white contour plots refer to the \textit{Chandra} emission below 0.8 keV, while the red ones the ratio shown in the right panel. \textit{Right:} \textit{Chandra} 0.8-1.3 to 0.3-0.8 keV ratio. Pixel corresponding to 0 counts in either image have been set to ratio 0. The red contour plots shown in left panel refer to this image, but the scale is different, as outlined by the broken-line rectangle. In both panels, the black cross indicates the position of the nucleus. See text for details.}
\end{figure*}

The optical image shows a prominent large-scale dust lane on the north-east. To better map its morphology, we divided the NIR image by the optical image. The result is shown in the left panel of Fig. \ref{dustmap}, where we also plotted the soft X-ray ($<0.8$ keV) contours, in white: the morphology of this emission is notably influenced by the dust lane. Most of the X-ray emission comes from the west side of the galaxy, which is less affected by absorption, but it is also present east of the nucleus, in the regions where the dust lane appears to be optically thinner. To better compare the optical and the X-ray obscuration, we extracted \textit{Chandra} images in the 0.8-1.3 and 0.3-0.8 keV bands. These bands were selected because their ratio is sensible to column densities of the order of  $10^{21}$ cm$^{-2}$. We are interested here in relative changes of the ratio between the two images. On the other hand, their absolute ratio has not straightforward physical interpretation, being dominated by the transfer function of the instrument, and will not be considered hereafter. Therefore, we normalized each image to its maximum, allowing us to achieve similar dynamical ranges in all images. Then, we built an image of the ratio of the two images. The result is shown in the right panel of Fig. \ref{dustmap} and by the red contours in the left panel. Both the region of higher ratio (high X-ray absorption) and the dust lane are located north-east of the nucleus.

Unfortunately, no [{O\,\textsc{iii}}] $\lambda 5007$ \textit{HST} image is available, thus preventing us to perform a detailed comparison between the soft X-ray emission and the NLR as done for other Seyfert 2s by  \citet{bianchi06}. However, the shape and dimension of the [{O\,\textsc{iii}}] cone as observed by the CTIO's 91-cm telescope closely resemble the soft X-ray contours plotted in Fig. \ref{dustmap} \citep{sbb91}.

\begin{figure*}
\begin{center}
\epsfig{file=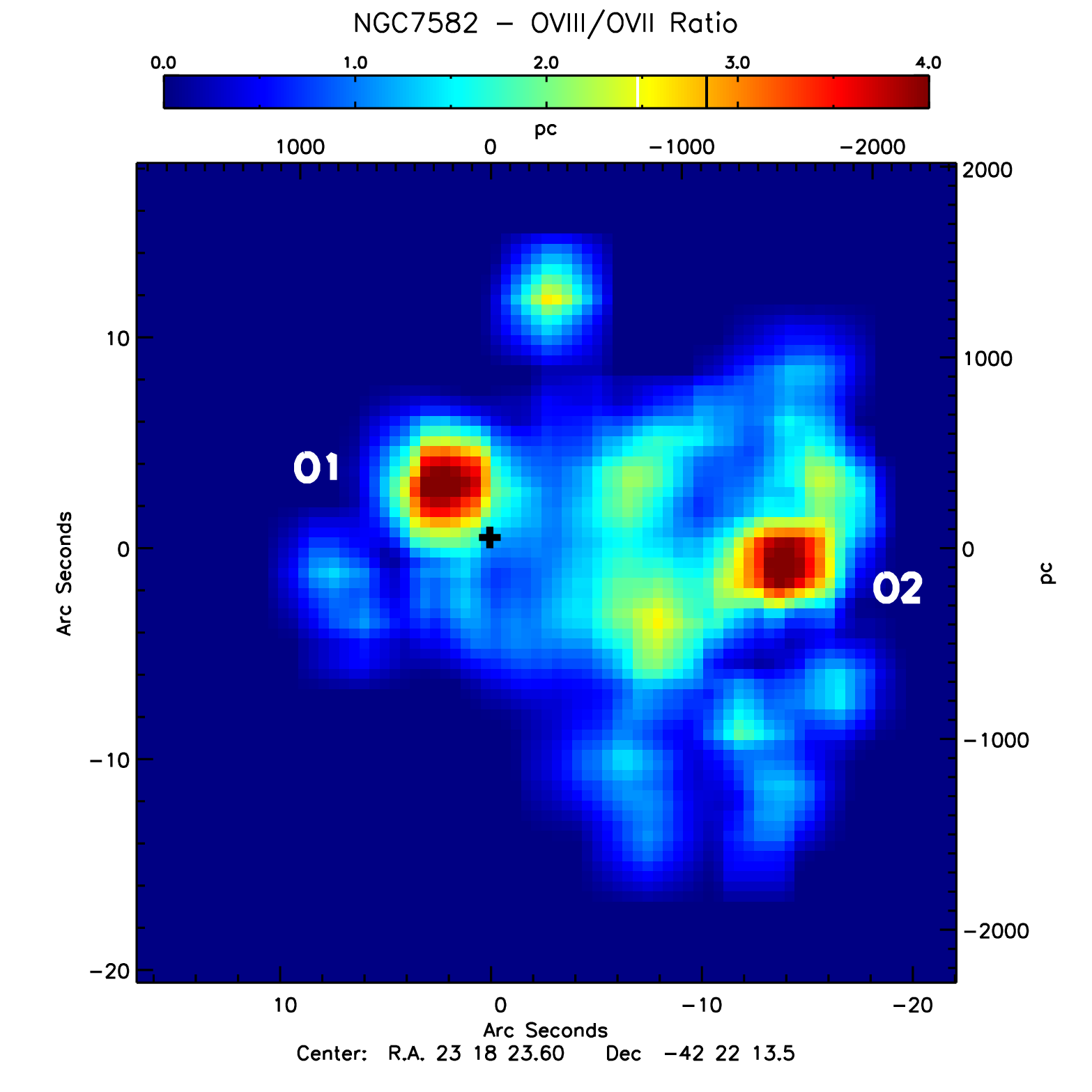, width=8.8cm}
\epsfig{file=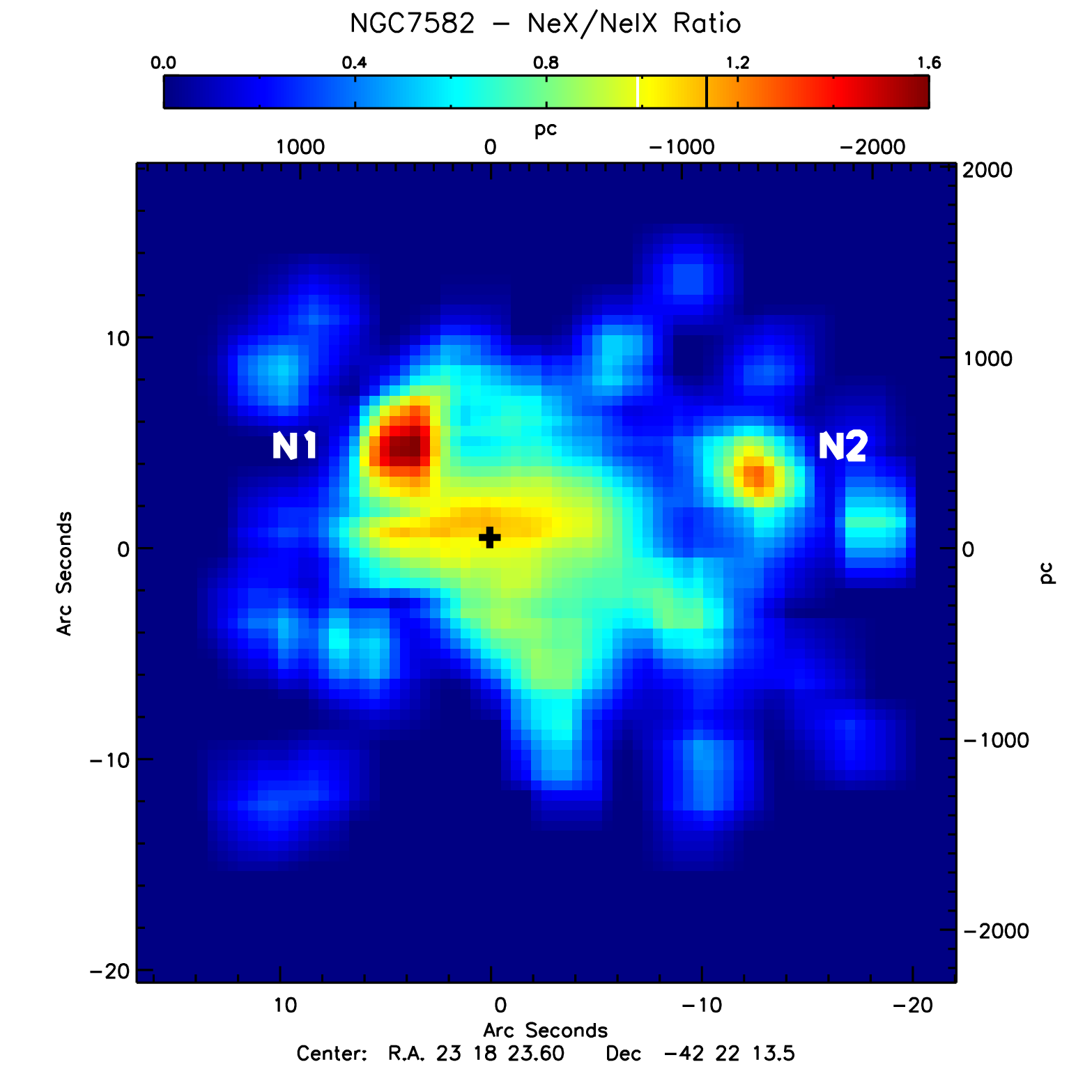, width=8.8cm}
\end{center}
\caption{\label{ratio}\textit{Left}: Ratio between the 0.6-0.7 and 0.5-0.6 keV images of NGC~7582, effectively mapping the {O\,\textsc{viii}} to {O\,\textsc{vii}} ratio along the soft X-ray emitting gas. \textit{Right}: same as above, but for the 0.95-1.1 keV and 0.85-0.95 images, mapping {Ne\,\textsc{x}} and {Ne\,\textsc{ix}}. Pixel corresponding to 0 counts in either image have been set to ratio 0. In both panels, the black cross indicates the position of the nucleus. See text for details.}
\end{figure*}

The XMM-\textit{Newton} RGS spectrum of NGC~7582 shows that the soft X-ray spectrum is dominated by a wealth of emission lines, with a very low level of continuum emission (Piconcelli et al., in preparation). Therefore, we extracted \textit{Chandra} images in narrow bands to look for spatial variations of emission lines' ratio. In particular, we selected 4 bands on the basis of the RGS spectrum, each one dominated by emission lines of a particular ion: 0.5-0.6 keV ({O\,\textsc{vii}}), 0.6-0.7 keV ({O\,\textsc{viii}}), 0.85-0.95 keV ({Ne\,\textsc{ix}}), 0.95-1.1 keV ({Ne\,\textsc{x}}). For each element, we then performed, as explained above, an image of the ratio of the higher to the lower ionization stage. The result is shown in Fig. \ref{ratio}: two `hot spots' are identified for oxygen (labelled O1 and O2) and for neon (N1 and N2). They point to regions where emission from the higher ionization stages is more intense. The significance of these spots (calculated on the ratio of the counts of the original, unsmoothed images) is 2$\sigma$ for O1, O2 and N1, while 1$\sigma$ for N2. The positions of the oxygen and the neon spots are not fully coincident, but they follow a very similar pattern. A possible explanation for this difference may be the presence of a not-negligible contamination by Fe L in the Neon dominated bands, possibly affecting the significance and location of N1 and N2.

\section{Discussion}
\label{discussion}
Spatially-resolved X-ray spectroscopy is beyond the capabilities of present missions, at least for source as bright and far as the brightest nearby AGN. Nonetheless, the unprecedented spatial resolution of the \textit{Chandra} optics permits energy-resolved imaging analysis, even when they are coupled with a low energy resolution detector. This technique has not been fully exploited yet \citep[but see some examples in, e.g,][]{yws01,iwa03,colb05}. Once combined with data from other wavelengths, it provides information on the physical states and the geometrical distribution of the circumnuclear gas, which cannot be obtained otherwise. We have shown in the previous section that NGC~7582 is one of the few nearby obscured AGN where energy-resolved imaging analysis is indeed possible with current instrumentation. In this section, we will discuss the results obtained in Sect \ref{analysis} in the context of an overall interpretative scenario for the nuclear region.

\subsection{\label{dust}Dust and X-ray absorption}

The left panel of Fig. \ref{dustmap} shows how the dust lane influences the morphology of the soft X-ray emission in NGC~7582. Most of the soft X-ray emission comes from the south-west, exactly as the optical [{O\,\textsc{iii}}] ionization cone, complementarily to the presence of optical obscuration. X-ray obscuration, therefore, appears to be coincident with the dust lane. This is confirmed by the right panel of Fig. \ref{dustmap}, which directly maps X-ray obscuration: the east side of the `cone` becomes visible at higher energies, below the thinner part of the dust lane (see also red contours in the left panel).

While it is not possible to be strictly quantitative in the comparison of the dust lane and the X-ray absorber with the present data, we can perform a simple, qualitative analysis.  Adopting the standard Galactic gas-to-dust ratio, the optical reddening may be rewritten in terms of the absorbing column density via the relation $A_V=5.27\,N^{22}_{H}$ mag, where the column density is expressed in units of $10^{22}$ cm$^{-2}$ \citep[see e.g.][and references therein]{mai01}. As an order of magnitude estimate, the column density required to obscure emission below 0.8 keV is of several $10^{21}$ cm$^{-2}$. Such column density would correspond to $A_V$ of the order of unity ($\simeq3$ mag for $N_H=5\times10^{21}$ cm$^{-2}$), which is roughly in agreement with the observed dust.

\subsection{\label{ionizationmap}The `hot spots'}

Fig. \ref{ratio} shows that the soft X-ray emitting gas in this source is not homogeneous. Two `hot spots', where {O\,\textsc{viii}} and {Ne\,\textsc{x}} emission are brighter than {O\,\textsc{vii}} and {Ne\,\textsc{ix}}, respectively, are present on both sides of the nucleus.  The origin of these `hot spots' may be at least twofold: i) variation of the obscuring column density or ii) variation of the ionization parameter of the gas (U). 

In the first case, the higher observed ratio would be an effect of absorption: the enhancement of the emission from higher ionization stages would be driven by the fact that it is less influenced by the intervening column density. This seems a reasonable hypothesis for O1/N1, which are indeed in the region covered by the thinner part of the dust lane (see Fig. \ref{dustmap}). The column density inferred for this region in the previous section, a few $10^{21}$ cm$^{-2}$, is indeed of the order of magnitude needed to affect the ratio between the emission lines under analysis. On the other hand, no clear absorption structures are present around O2/N2, so that this explanation seems inappropriate for these `hot spots'.

In the second scenario, the variation of the lines ratio would be a tracer of spatial heterogeneity of U. An higher value of U may be due to a decrease of the density or a change of the ionizing flux, as in the case of the presence of a localized photoionizing source. \citet{bianchi06} have shown that a gas with constant U can reproduce the average NLR/soft X-ray properties. Assuming the central AGN as the only source of photoionization, a constant U at every radius implies that the density should be falling away from the nucleus like $r^{-2}$. Any local deviation from this density law would produce spots of higher or lower ionization, as the ones observed in NGC~7582.

On the other hand, the value of U can be locally enhanced by a photoionization source other than the central AGN, like radio shocks or star-formation regions. Indeed, it is interesting to note that NGC~7582, differently from the Seyfert 2s presented by \citet{bianchi06}, is known to host a strong starburst  \citep[see e.g.][]{wg06}. The RGS spectrum of NGC~7582 is indeed dominated by emission lines, as many Seyfert 2s, but the strongest line is the {O\,\textsc{viii}} K$\alpha$, as often found in strong starburst galaxies (Guainazzi \& Bianchi, in preparation). However, the {O\,\textsc{viii}} `hot spots' are very far from the strong star-formation regions resolved in the optical and NIR images (see Fig. \ref{nucleus}) and, therefore, a direct relation between the two phenomena seems rather unlikely. Moreover, the soft X-ray emission is much more extended than the starburst region and not obviously correlated, as it is with the NLR. Therefore, even if we cannot exclude that the continuum emission produced by the starburst contributes to the photoionization of the NLR, it seems that it is not the principal source of ionizing photons and cannot account for the observed `hot spots'.

As far as we know, this is the first evidence presented so far for regions with different spectral properties in the soft X-ray emitting gas of an AGN. Whether this is found to be common in Seyfert 2s or not in the future, it will provide unique information to better understand the physical processes at work in the extended soft X-ray emission of AGN and its relation to the materials observed in other bands.

\subsection{\label{BLR}The geometry of the absorbers}

The joint analysis of \textit{Chandra} and \textit{HST} data of NGC~7582 requires the presence of three absorbers. First, the near infrared image is dominated by a central, unresolved, source, which, as generally believed in Seyfert 2s, is likely to be re-emission of the central AGN from the dusty torus. However, the nuclear optical emission is a factor $\simeq1000$ lower. This ratio between the fluxes at 1.6 and 0.6 $\mu$m is by far the largest if compared with a sample of Seyfert 2 nuclei observed by \textit{HST} \citep[e.g.][]{alherr03}, thus being suggestive of a further obscuration of the torus itself by a second intervening material. 
Moreover, a significant part of the soft X-ray emission is strongly affected by large-scale absorption, correlated to the dust lane, leaving one side of the cone completely unabsorbed. These results are in agreement with the XMM-\textit{Newton} spectrum, which is absorbed by at least two materials, the thickest being likely the torus (Piconcelli et al., in preparation). Note that neither the RGS nor the EPIC pn spectra, extracted from much larger regions, can measure the large-scale absorption in the soft X-ray emission, because it only obscures a part of the emission. 

This scenario fits well with the results presented by \citet{gua05b} on a large sample of Seyfert 2s. They found a significant correlation of X-ray obscuration in the range $10^{22}-10^{23}$ cm$^{-2}$ with the presence of $\simeq100$-pc scale nuclear dust. Therefore, as proposed by \citet{malk98} and \citet{matt00b}, a possible extension of the standard unified model should include, as well as the Compton-thick torus, a Compton-thin material located on a larger scale, associated with dust lanes. As already noted by \citet{gua01}, this material is not necessarily coincident with the dust lane, but a ``dusty environment'' could favour the formation of large scale X-ray absorbing clouds. In particular, in the case of NGC~7582, we suggest that the dust lane may be directly responsible for the absorption of the north-east part of the soft X-ray emission, while a thicker cloud is along the line of sight to the torus.

\section{Conclusions}

A multiwavelength approach to imaging analysis of AGN is still underutilized, in particular with X-ray data, but it provides a unique opportunity to study the geometry of the circumnuclear materials and their mutual relations. With this aim, we have performed a detailed analysis of the \textit{Chandra} and \textit{HST} images of NGC~7582, whose results can be summarized as follows:

-The dust lane appears to be coincident with a region where the soft X-ray emission is absorbed. Under standard assumptions, the derived gas column density and dust $A_V$ are roughly in agreement. Interestingly, soft X-ray emission at higher energy becomes visible in regions where the dust lane appears thinner.

-Two `hot spots', i.e. regions where emission from higher ionization stages of O and Ne is stronger, are observed in the soft X-ray emission. While in one case the effect due to absorption may be the most important one, these `hot spots' may be in principle tracers of heterogeneity of the emitting gas. Moreover, the lack of any coincidence between the star-formation regions (as observed in the \textit{HST} optical and NIR images) and the `hot spots'  seems to exclude that, at least in NGC~7582, the starburst has an important role as a source of photoionization.

-The overall picture is consistent with unified scenarios where a compact torus intercepts the line of sight only in Compton-thick AGN, while a large-scale Compton-thin material is associated with the host galaxy matter on larger scale. 

\section*{Acknowledgements}

SB would like to thank the STScI Visitor Program for the funding and the hospitality. We acknowledge an anonymous referee for valuable comments.

\bibliographystyle{mn}
\bibliography{sbs}

\label{lastpage}

\end{document}